\documentclass[conference]{IEEEtran}
\usepackage{blindtext, graphicx, tabularx, booktabs, amsmath, siunitx, subfigure, xcolor}

\usepackage{cite}

\usepackage{color}
\usepackage{soul}
\usepackage{flushend}

\usepackage{algorithm} 
\usepackage{algpseudocode}

\newcolumntype{R}[1]{>{\raggedright\arraybackslash}p{#1}}

\IEEEoverridecommandlockouts

\begin{document}

\title{


Improving Content-Aware Video Streaming in Congested Networks with In-Network Computing


\thanks{This work was financed in part by the Coordena\c{c}\~{a}o de Aperfei\c{c}oamento de Pessoal de N\'{i}vel Superior - Brasil (CAPES) - Finance Code 001, CNPq, and FAPERGS. This work also received funding from Grant \#2020/05183-0, São Paulo Research Foundation (FAPESP) and National Science Foundation (NSF) Award \#1740791.}

}

 \author{
     \IEEEauthorblockN{
        Leonardo Gobatto\IEEEauthorrefmark{1},
        Mateus Saquetti\IEEEauthorrefmark{1},
        Claudio Diniz\IEEEauthorrefmark{1},
        Bruno Zatt\IEEEauthorrefmark{2},
        Weverton Cordeiro\IEEEauthorrefmark{1},
        Jose R. Azambuja\IEEEauthorrefmark{1}
     }
  
      \IEEEauthorblockA{
          \IEEEauthorrefmark{1}Federal University of Rio Grande do Sul (UFRGS) - Institute of Informatics - PPGC - PGMICRO
          \\\{leonardo.gobatto, mateus.saquetti, claudio.diniz, weverton.cordeiro, jose.azambuja\}@inf.ufrgs.br
      }
    
      \IEEEauthorblockA{
          \IEEEauthorrefmark{2}Federal University of Pelotas (UFPEL) - CDTEC - PPGC
          \\zatt@inf.ufpel.edu.br
      }

}

\maketitle

\begin{abstract}

Network congestion and packet loss pose an ever-increasing challenge to video streaming. Despite the research efforts toward making video encoding schemes resilient to lossy network conditions, forwarding devices have not considered monitoring packet content to prioritize packets and minimize the impact of packet loss on video transmission. In this work, we advocate in favor of in-network computing employing a packet drop algorithm and an in-network hardware module to devise a solution for improving content-aware video streaming in congested network. Results show that our approach can reduce intra-predicted packet loss by over 80\% at negligible resource usage and performance costs.

\end{abstract}
    

\section{Introduction}
\label{sec_introduction}

Video streaming has been the main driving force of Internet traffic growth in the past few years, having accounted for over 60\% of the global traffic in 2019~\cite{sandvine_2019} and almost 50\% of mobile traffic in 2021~\cite{sandvine_2021}. Many reports also predict that streaming will jump to 80\% of global traffic share in the coming years~\cite{sandvine_2019}. This trend is explained by three main factors: (\emph{i}) the higher number of connected multimedia devices, estimated to be more than three times the global population by 2023~\cite{cisco_2021}; (\emph{ii}) the increased use of high and ultra-high resolution videos, 
e.g., 66\% of connected flat-panel TVs will support 4K ($3840 \times 2160$ \textit{pixels}) by 2023~\cite{cisco_2021}; and (\emph{iii}) the popularization of video streaming due to video-on-demand (VoD) services.

The growing demand for video streaming will pose ever-increasing stress on the global networking infrastructure, a situation only worsened with the coronavirus pandemic~\cite{nytimes2020}. It means that, to keep up with users' expectations for high-quality streaming, networking researchers and practitioners will need to upgrade the networking infrastructure (e.g., with faster links) and devise solutions to optimize its usage (encoding schemes, delivery strategies, etc.). In this paper, we focus on delivering strategies to optimize video streaming network usage.

One such strategy is processing video streaming packets as they transit the network \cite{in-network-view-synthesis,10.1145/3362068}, by taking into account current network conditions (e.g., the proportion of users from a given network region interested in the streaming, available network bandwidth, etc.), thus complementing the decades-old rate-limiting video transmission strategy based on the end-user software feedback \cite{cowan1995adaptive}. Nonetheless, network congestion and packet losses remain a significant challenge, and encoding schemes resilient to losses become paramount.

In this context, the bitstream in recent video coding standards such as H.264/AVC (Advanced Video Coding)~\cite{H264} and H.265/HEVC (High-Efficiency Video Coding)~\cite{H265} is partitioned into Network Abstraction Layer (NAL) units to facilitate video transmission over lossy packet-based networks~\cite{Sjoberg:2012}. Unlike previous standards, H.264/AVC and H.265/HEVC are robust to packet losses since video can be decoded even when some packets are lost. This is done with the division of coded video data and metadata into NAL units of different types, Intra Random Access Point (IRAP) and non-IRAP picture NAL units. The former contains video data encoded with intra prediction and is self-contained since it does not need other NAL units to be reconstructed. IRAP NAL units are particularly important in the decoding process, as they are used as a reference to reconstruct non-IRAP pictures that employ inter prediction through motion compensation.

Due to the error propagation in frame encoding, packet losses influence video quality more aggressively when IRAP NAL units are lost in the error-prone packet-based networks. To the best of our knowledge, no solution has explored the idea of discarding non-IRAP packets under network congestion. One option to address the network congestion issue is to devise an in-network mechanism capable of monitoring the video stream and selectively discarding packets that pose less impact to the video quality. Recent advances in network programmability and computer architectures have allowed designers to move the computation inside the network, where it is closest to the data \cite{daCostaCordeiro2017,sapio2017network}. Even though we cannot stop packet loss, we can analyze packets going through the network and prioritize data, such as IRAP over non-IRAP NAL units.

To bridge this gap, we propose an in-network computing approach for selective video streaming packet discarding under network congestion. In our proposal, we monitor video streaming based on video coding standards such as H.264/AVC and H.265/HEVC and selectively discard non-IRAP picture NAL unit packets in a content-aware network flow. To this end, we rely on a hardware architecture that extends traditional packet-forwarding devices (e.g., switch) employing network programmability to analyze network flows, detect congestion, and preemptively discard non-IRAP NAL units. We implemented our algorithm as an in-network computing module to a layer-2 switch and deployed it on a NetFGPA-SUME board. We evaluated our solution by implementing a congestion simulator that emulates a switch with configurable packet input and output ratios, packet buffer size, and congestion duration. Our findings show that with as little as 2\% additional hardware resources, one can reduce IRAP frames losses to an average of 4.8\%, an 82.5\% reduction against the baseline. These findings can help designers better understand the impacts of network congestion in video streaming and improve the quality of experience perceived by end users.

In summary, we make the following contributions:
(1) A content-aware algorithm for effectively dropping non-IRAP NAL units in congested networks; (2) An in-network computing hardware module to extend packet-forwarding devices; and (3) A simulation platform that emulates a packet-forwarding device in congestion environments.


\section{Background and Related Work}
\label{sec_background}

Video is transmitted over the Internet in compressed form, often using international standard codecs developed by ISO/IEC Motion Picture Experts Group (MPEG) and ITU-T Video Coding Experts Group (VCEG). These standards define the decoding process and the syntax of the compressed video (bitstream). Real-time video transmission over the Internet in the presence of packet losses is not a new problem, and many techniques were proposed to make it more resilient to errors~\cite{Wang:2000}. An important development in this subject was introduced in H.264/AVC standard~\cite{H264}. This standard was developed considering video transmission through packet-switching networks. The bitstream in H.264/AVC is partitioned into NAL units, separating video data and metadata and exposing some information to the transport layer. This is done through the NAL unit header, which includes a NAL unit type field to identify sequence and picture parameter sets (metadata information about video sequences and pictures), IDR pictures (Instantaneous  Decoding Refresh - used for random access in the bitstream), non-IDR pictures, and so on. The H.265/HEVC standard~\cite{H265} inherits this concept whereas defining the more general IRAP pictures/NAL~\cite{Sjoberg:2012}. 

Some works have evaluated the effects of packet losses in H.264/AVC and H.265/HEVC video transmission. The authors from~\cite{Korhonen2018} used H.264/AVC to present a subjective video degradation analysis under packet loss scenarios whereas differentiating the impact of reference and non-reference pictures. 
In~\cite{Kazemi2018}, an encoding-time technique is presented for defining intra-predicted frames (i.e., IRAP) positioning in the H.264/AVC bitstream to improve packet loss resiliency.
A similar strategy is adopted in~\cite{van2021} where IRAP frames (referred to as keyframes by the authors) are periodically inserted during the encoding process to improve H.265/HEVC bitstreams resiliency.
The work in~\cite{Oztas:2012} concluded that HEVC is less error resilient than H.264/AVC. The work in~\cite{Nightingale:2012} developed a framework for streaming and evaluation of video encoded with HEVC in lossy networks.

In the context of computer networks, many advances have occurred in the past decade, especially in terms of Software-Defined Networking (SDN). Such advances have led to the emergence of network programmability~\cite{rmt-sigcomm-2013,daCostaCordeiro2017} and have provided network administrators with the ability to reprogram the behavior of forwarding devices through Domain-Specific Languages (DSL) such as P4, POF, and Lyra~\cite{Bosshart:2014:P4,Song:2013:PFU:2491185.2491190,Lyra-sigcomm2020,npl}. In the same way that the networking infrastructure advanced in programmability, it also advanced in computational power. With new programmable network hardware and accelerators on the market, such as Smart Network Interface Cards (SmartNICs)~\cite{sanvito2018can}, Graphics Processing Units (GPUs)~\cite{sun2019optimizing}, and FPGAs~\cite{woodruff2019p4dns}, a new generation of programmable network devices is flooding the market, enabling computation to be performed within the network. However, in-network computing is still in its infancy, with most works targeting networking-related applications, such as caching~\cite{infocom_cache}, storage~\cite{tokusashi2018lake}, and data aggregation~\cite{sapio2017network}, with only a few recent works aiming at other areas such as artificial intelligence~\cite{li2019accelerating,comml_in-net} and adaptive video rates~\cite{ICNP_video,TransMulti_video}.
To the best of our knowledge, this is the first work in the literature to design and evaluate an in-network computing architecture to improve video streaming in congested networks at negligible costs in resource usage and performance degradation.

\section{Proposed In-Network Computing Architecture}
\label{sec_proposal}

The proposed in-network computing architecture discussed in this work targets a generic packet-forwarding device. It aims at decreasing IRAP packet loss during network congestion while maintaining packet loss to a minimum. To do so, we devised a content-aware algorithm for preemptively dropping non-IRAP packets and an in-network hardware module to prove the feasibility of our approach in terms of resource usage and performance. In the following, we describe in detail the packet drop algorithm and our in-network hardware module.

\subsection{Proposed Packet Drop Algorithm}

The main objective of our proposed packet drop algorithm is to avoid IRAP-packet loss. One option would be to detect network congestion and preemptively drop any non-IRAP packets until the congestion is over. However, this could lead to unnecessary non-IRAP packet loss, as the buffer may be able to store all IRAP packets forwarded during congestion and additional non-IRAP packets. Thus, a better option would be to evaluate how many packets we can store during the congestion based on packet throughput and buffer occupation. By doing so, we can guarantee that our proposed packet drop algorithm will not unnecessarily drop non-IRAP packets.

Algorithm~\ref{alg:DROP} describes the pseudocode of our proposed packet drop function. The algorithm starts when a congestion notification informs the beginning of congestion ($c\_start$) and its duration ($c\_time$). These data are used to start a timer ($timer$) that puts the system into congestion mode. While in congestion ($timer > 0$), we calculate the number of packets that will be received during the congestion ($packets$) by multiplying the packet throughput ($p\_tp$) by the remaining congestion time ($timer$). In case the buffer cannot hold all remaining packets that will be received during congestion ($b\_free < packets$) and the packet is a non-IRAP ($! p\_irap$), we recommend a packet drop ($return~true$). Otherwise, we recommend not dropping the packet ($return~false$). Note that this is only a recommendation to the packet-forwarding device, as the decision to drop packets is the forwarding device's.

\begin{algorithm}
    \scriptsize
	\caption{Drop Packet Policy}
	\label{alg:DROP}
	\begin{algorithmic}[1]
	    \Function{drop}{$c\_start,c\_time,b\_free,p\_tp,p\_irap$}
	        \If{$c\_start$}
	            \State $timer \gets c\_time$
            \EndIf	   
		    \If {$timer > 0$}
		        \State{$timer \gets timer - 1$}
		        \State $packets \gets p\_tp \times timer$
		    \If{$b\_free < packets \And ! p\_irap$}
		        \State \Return {$true$}
		    \EndIf
		    \EndIf
		    \State \Return {$false$}
	    \EndFunction
	\end{algorithmic} 
\end{algorithm}

\subsection{Proposed In-Network Hardware Module}

\begin{figure}[!tb]%
    	\centering
        \includegraphics[width=1\columnwidth]{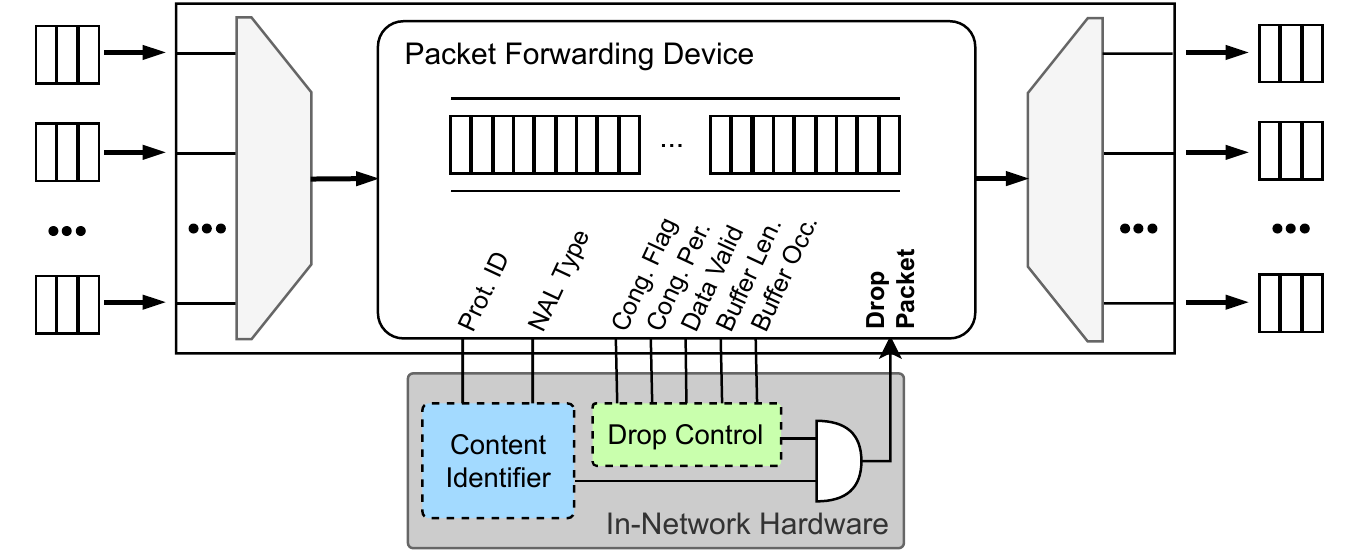}
        \caption{In-network Hardware Module Diagram}
        \label{fig:in-network_HW}
    \end{figure}

To evaluate resource usage and performance, we developed an in-network hardware module that extends a generic packet-forwarding device through a simple interface. As shown in Fig.~\ref{fig:in-network_HW}, the in-network hardware implements our proposed algorithm with two blocks, the Content Identifier and the Drop Control. The Content Identifier receives a protocol identifier (\textit{Prot. ID}) and the NAL type (\textit{NAL Type}). It then parses this information and returns true if it detects a non-IRAD packet. The Drop Control block inputs information on the congestion (\textit{Cong. Flag} and \textit{Cong. Per.}), the validity of the data received (\textit{Data Valid}), and information on the buffer status (\textit{Buffer Len.} and \textit{Buffer Occ.}), returning true if there is not enough space left in the buffer until the end of the network congestion. When both these flags are true, the in-network hardware module indicates to the packet-forwarding device that the packet should be dropped.

We implemented a layer-2 switch (l2-switch) as a baseline packet-forwarding device in P4~\cite{Bosshart:2014:P4} and our proposed in-network hardware module in Verilog. To interface both modules, we modified the l2-switch pipeline (in P4) to detect NAL headers and forward them through an external module to our hardware module (in Verilog). We then synthesized both projects (l2-switch and l2-switch extended) to the NetFPGA-SUME board with the Xilinx SDNet high-level synthesis tool, following the P4VBox~\cite{p4vbox} workflow. Our resource usage and performance evaluations were performed on Xilinx Vivado 2018.2 by injecting custom NAL packets.

Table~\ref{tab:hw_occupation} shows resource usage and performance of a case-study l2-switch and the same l2-switch with our extended in-network hardware module for packet dropping obtained in Xilinx Vivado for a NetFPGA-SUME. In terms of resource usage, our implementation required additional 2.6\% LUTs and 1.9\% FFs. Note that the l2-switch is one of the simplest packet-forwarding devices, and our implementation is agnostic to the forwarding device. Therefore, more complex devices would perceive an even smaller impact on resource usage. Considering performance, the extended l2-switch achieved the same latency and throughput as the baseline, without degradation.

\begin{table}[h]
\centering
\caption{In-network hardware occupation}
\begin{tabular}{l|r|r|r}
\toprule
Resource and Performance & L2-switch & L2-switch extended & Overhead (\%) \\
\midrule
LUTs                & 1,774  & 1,821    & 47 (2.6\%) \\
FFs                 & 3,831  & 3,904    & 73 (1.9\%) \\
Latency ($\mu s$)   & 0.56   & 0.56     & -- \\
Throughput ($Gbps$) & 101.1  & 101.1    & -- \\
\bottomrule
\end{tabular}
\label{tab:hw_occupation}
\end{table}

\section{Evaluation}
\label{sec_evaluation}

Our evaluation consists of transmitting a set of videos through an emulated packet-forwarding device under congestion environments. To do so, we developed a packet-forwarding device simulator supporting tools to analyze an H.265/HEVC bitstream flow in congestion situations. The simulator consists of a configurable test bench, where users can adjust packet input and output rates, buffer size, and congestion frequency and duration (during congestion, output rate drops to zero). Our supporting tools adapted the tools available at~\cite{h26x_extractor} and expanded them to work with H.265/HEVC bitstreams, parsing relevant information on the NAL units, such as the number of NAL units, packets per NAL, bytes per NAL, percentages of NAL types, among others.

For this work, we configured our simulator with a buffer size of 60 packets, a realistic value considering the target board used for synthesizing our proposed in-network hardware module. Considering that forwarding packet devices usually operate at line rate, having buffers specifically for cases of network instability, and for simplicity, we set the packet input and output rates at 60 and 120 packets per time unit, respectively. By doing so, the buffer will start filling as soon as the congestion starts and will be cleared as soon as the congestion ends. Additionally, we can manage packet loss by adjusting the frequency and duration of the congestion scenarios. To adjust our experiments for packet losses from 5\% to 50\%, in a 5\% step, we varied congestion times from 10\% to 94.11\% of the simulation time.

As case-study video benchmark, we used all 24 videos from the Common Test Conditions (CTC) of H.265/HEVC ~\cite{CTC:2013} encoded using x265 with the four recommended Quantization Parameters (QP - 22, 27, 32, and 37). The CTC contains videos with multiple resolutions ($416 \times 240$, $832 \times 480$, $1024 \times 768$, $1280 \times 720$, $1920 \times 1080$, and $2560 \times 1600$), including different motion and texture characteristics. Considering that the videos are relatively short, we chose to concatenate them 100 times before sending them through our simulator, to reach statistical relevance ($\sigma = 0.0068$).

\begin{figure}[!tb]%
    	\centering
        \includegraphics[width=0.85\columnwidth]{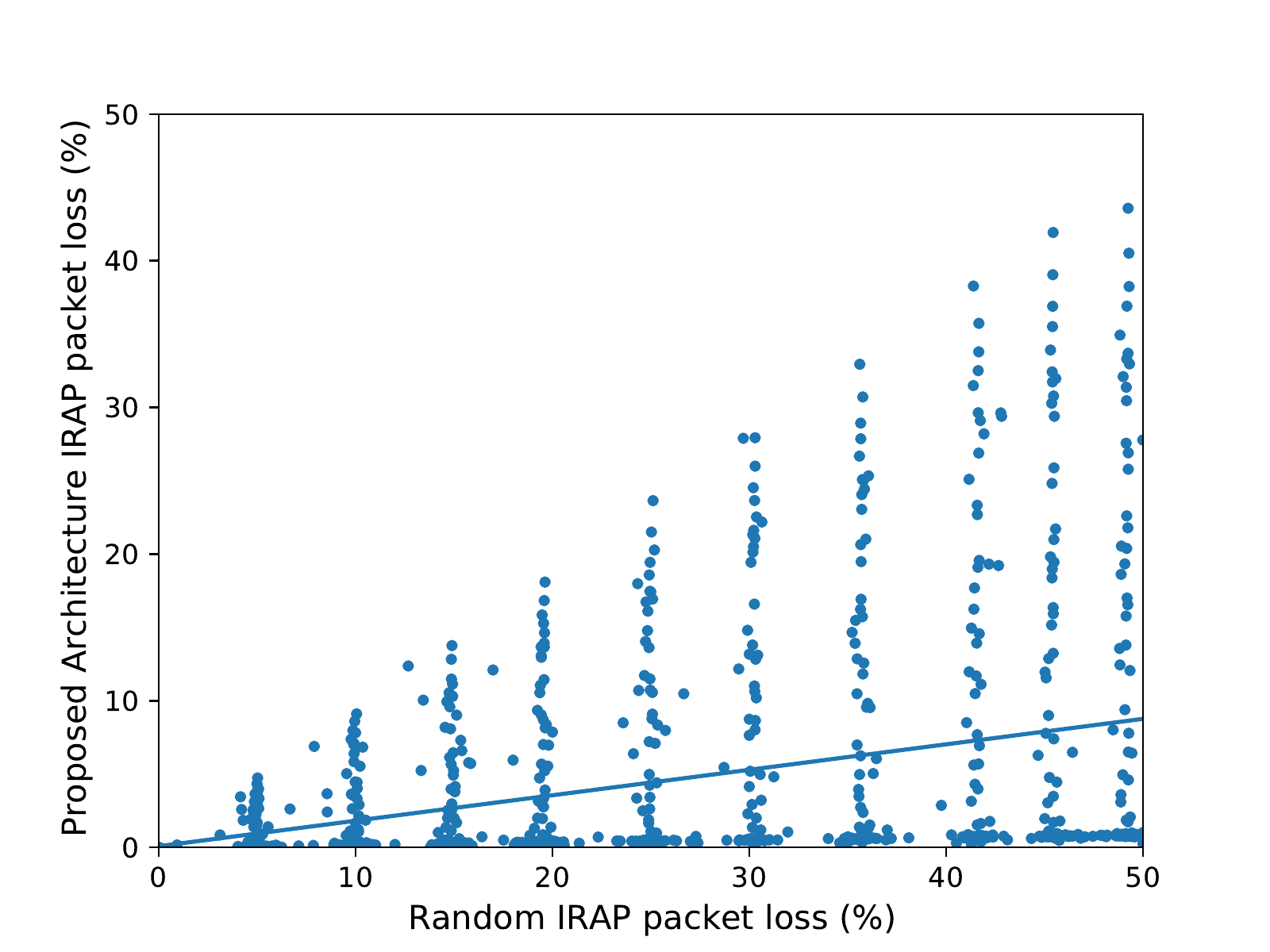}
        \caption{IRAP packet loss}
        \label{fig:2}
    \end{figure}

Fig.~\ref{fig:2} correlates the random IRAP packet loss (X-axis), simulating the baseline l2-switch, with our proposed architecture's IRAP packet loss (Y-axis), simulating the l2-switch extended, for all 24 benchmark videos encoded with 4 QPs and 10 packet loss rates, in a total of 960 simulations. As one can notice, all points are sub-linear ($y < x$), showing that our approach is never worse than the random one. Instead, the drawn line shows the linear regression with an average IRAP packet loss of 4.8\%, 82.5\% lower than the average 27.5\%. However, even though the linear regression line is below 10\%, some points presented worse results.

\begin{figure}[!tb]%
    	\centering
        \includegraphics[width=0.85\columnwidth]{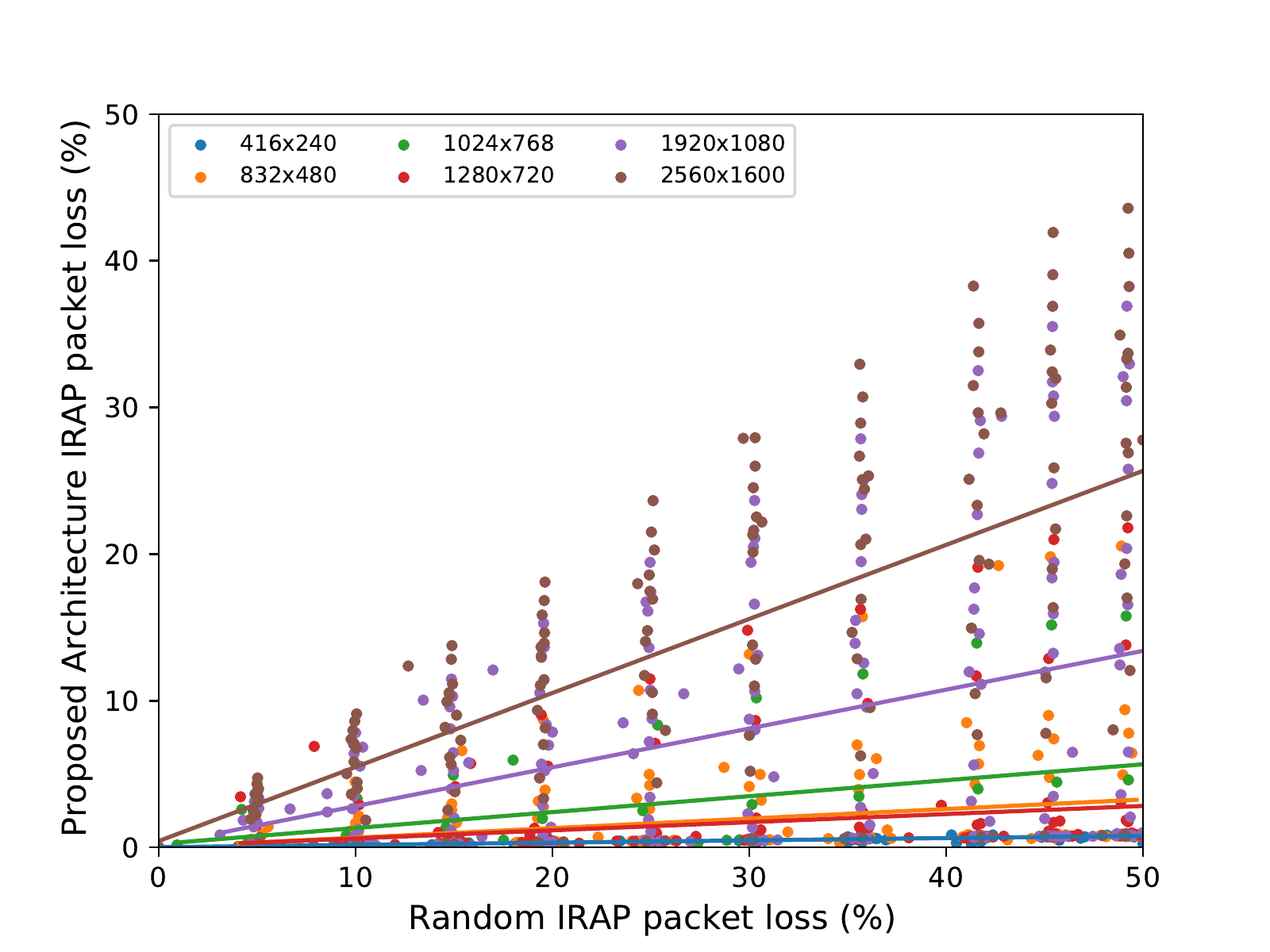}
        \caption{IRAP packet loss grouped by resolution}
        \label{fig:3}
    \end{figure}

Fig.~\ref{fig:3} shows the same data as Fig.~\ref{fig:2} but colored according to video resolution. It makes clearer that our solutions, on average, deals better with lower-resolution videos ($416 \times 240$, $832 \times 480$, $1024 \times 768$, and $1280 \times 720$) than with higher-resolution ones ($1920 \times 1080$ and $2560 \times 1600$). Considering the linear regression lines, the $2560 \times 1600$ videos show 33 times more IRAP packets lost than the $416 \times 240$. This happens mainly because of the ratio between the number of packets needed to send an IRAP NAL and the number of packets that the packet-forwarding device can hold in its buffer. Considering that packets will be lost during congestion, higher resolution videos have a higher chance of sending a single frame during the time of congestion, thus decreasing our proposed approach's gain over the baseline. However, the plot also shows that, even for high-resolution videos, results vary in the Y-axis, showing a margin to improve results.

\begin{figure}[!tb]%
    \centering
    \subfigure[1920x1080 ]{
        \label{fig:4a}
        \includegraphics[width=0.85\columnwidth]{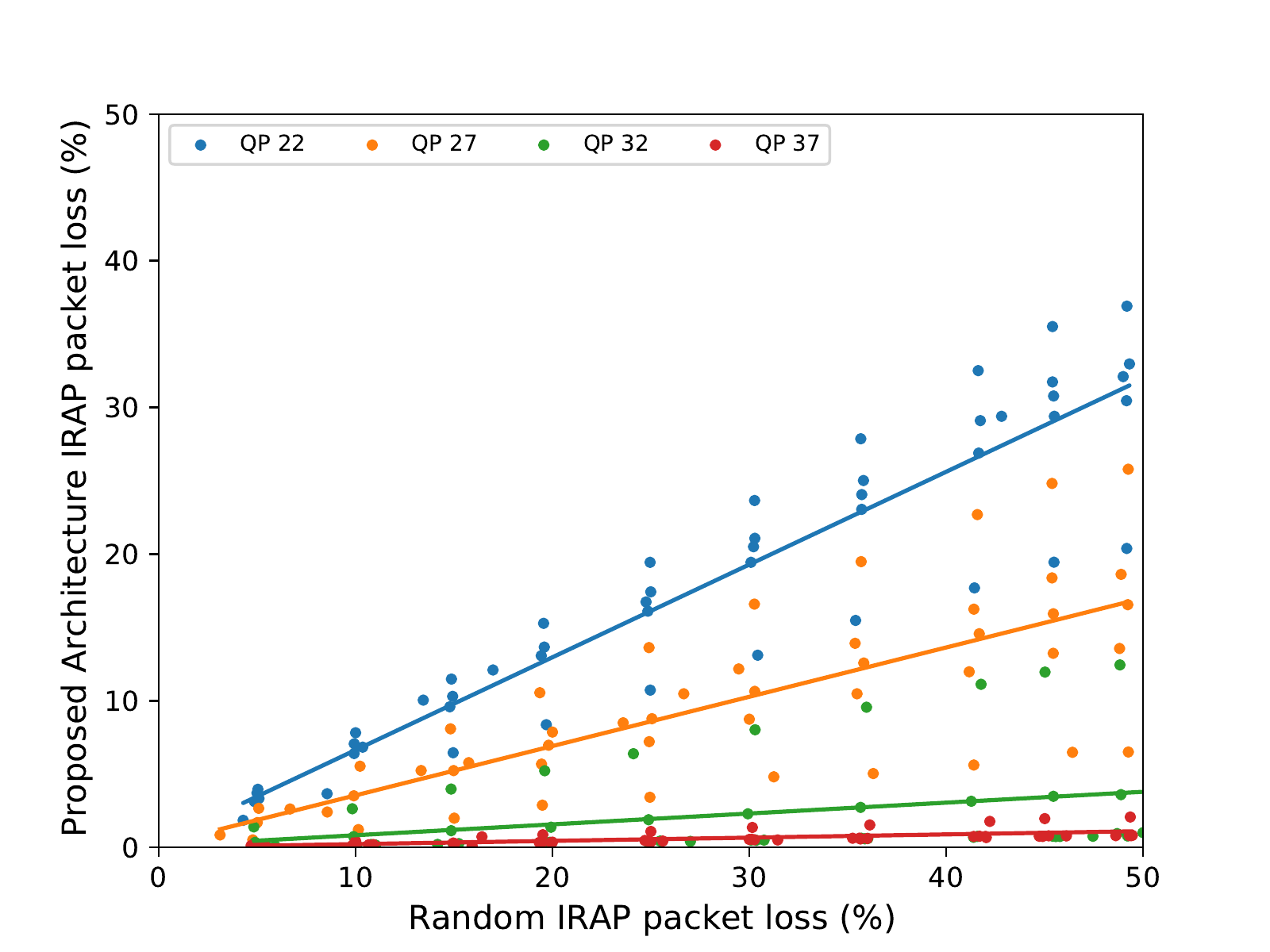}
    }%
    \qquad
    \subfigure[2560x1600]{
        \label{fig:4b}
        \includegraphics[width=0.85\columnwidth]{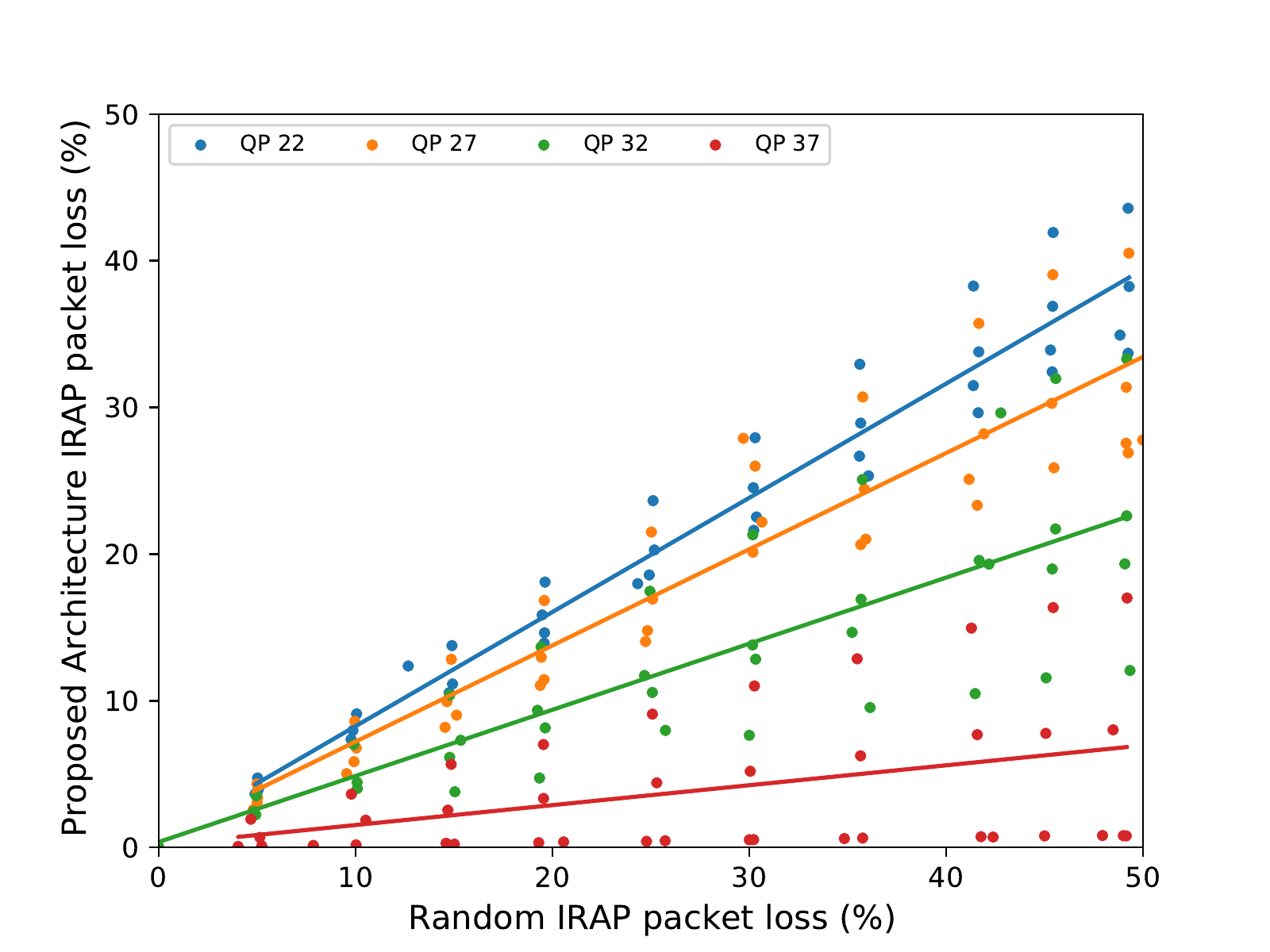}
    }%
    \caption{IRAP packet loss grouped by QP.}
    \label{fig:4}
\end{figure}

Fig.~\ref{fig:4} shows the data points constrained to the higher-resolution videos at $1920 \times 1080$ (Fig.~\ref{fig:4a}) and $2560 \times 1600$ (Fig.~\ref{fig:4b}), thus showing the impact of different QPs. As one can notice, the QP affects our proposed architecture when dealing with higher-resolution videos. For the $1920 \times 1080$ resolution videos, we can observe IRAP packet loss reduction, on average, in 96\% by increasing QP from 22 to 37. For the $2560 \times 1600$ resolution videos, we can reduce packets lost, on average, from 21.9\% to 3.9\% (82\% reduction).
Overall, our solution provides higher improvements over the random solution for higher QP values. Similar to what happens for high-resolution videos (recall Fig.~\ref{fig:3}), low QP values lead to larger bitstreams and, as a result, to a higher number of packets per IRAP picture. Thus, the ratio of packets to transmit an IRAP picture per number of packets in the buffer grows higher and limits the benefits of our solution.

Combined, these results show that our approach can reduce IRAP packets lost during network congestion at a negligible cost in resource usage and performance degradation. They also show that our proposed architecture works seamlessly for lower-resolution videos but requires attention when dealing with higher-resolution videos, especially when considering low QPs. Additionally, we believe that it is possible to improve results for packet-forwarding devices with larger packet buffers.

\section{Final Considerations}
\label{sec_conclusion}

This work tackled the issue of the ever-increasing stress on the global network infrastructure due to video streaming with the recent technology of in-network computing. We presented an algorithm for prioritizing IRAP packets in network congestion situations agnostic to the packet-forwarding device, an in-network hardware module that can be deployed to FPGA boards, and a simulator to evaluate congested environments. Our comprehensive H.265/HEVC video benchmark results show that our solution can decrease IRAP packet loss by over 82\% with negligible cost in resource usage and performance. As future work, we intend to further leverage content awareness by considering distinct drop priorities for non-IRAP pictures according to their hierarchy in the group of pictures and experiment with different buffer sizes and transmission technologies, such as HLS and MPEG-DASH.

\ifCLASSOPTIONcaptionsoff
  \newpage
\fi

\bibliographystyle{IEEEtran}
\bibliography{ref}

\end{document}